\documentstyle[12pt]{article}

\title{A solvable model of a random spin-$\frac{1}{2}$ $XY$ chain}
\author{
     Oleg Derzhko\\
{\em {Institute for Condensed Matter Physics}}\\
{\em {1 Svientsitskii St., L'viv-11, 290011, Ukraine}}\\
[5pt]
     Johannes Richter\\
{\em {Institut f\"{u}r Theoretische Physik,
      Universit\"{a}t Magdeburg}}\\
{\em {P.O.Box 4120, D-39016, Magdeburg, Germany}}}                           
\date{\today}
 
\begin{document}

\maketitle

\begin{abstract}
\normalsize
The paper presents 
exact calculations of thermodynamic quantities 
for the spin-$\frac{1}{2}$ isotropic $XY$ chain 
with random lorentzian intersite interaction 
and transverse field that 
depends linearly on the surrounding intersite interactions.

\end{abstract}

\vspace{2mm}

\noindent
{\bf PACS codes:}
{\em 75.10.-b}

\vspace{1mm}

\noindent
{\bf Keywords:}
{\em
spin-$\frac{1}{2}$ $XY$ chain,
lorentzian disorder,
Green functions approach,
density of states,
thermodynamics,
magnetization,
susceptibility}\\

\vspace{2mm}
\noindent
{\bf Postal addresses:}\\

\vspace{1mm}
\noindent
{\em 
Dr. Oleg Derzhko (corresponding author)\\
Institute for Condensed Matter Physics\\
1 Svientsitskii St., L'viv-11, 290011, Ukraine\\
Tel: (0322) 427439\\
Fax: (0322) 761978\\
E-mail: derzhko@icmp.lviv.ua\\

\vspace{1mm}
\noindent
Prof. Dr. Johannes Richter\\
Institut f\"{u}r Theoretische Physik, Universit\"{a}t Magdeburg\\
P.O.Box 4120, D-39016, Magdeburg, Germany\\
Tel: 0049 391 671 8670\\
Fax: 0049 391 671 1131\\
E-mail: johannes.richter@physik.uni-magdeburg.de}

\clearpage

\section{Introduction}

Starting from the seminal paper by E.Lieb, T.Schultz and D.Mattis 
the study of the one-di\-men\-sional 
spin-$\frac{1}{2}$ $XY$ models has attracted much
interest. A lot of exact results concerning
thermodynamics, spin correlations and their 
dynamics were found over the last 35 years.$^{1-6}$
The analytical
results obtained for random versions of such models are not so
impressive. 
Some results 
dealing with magnetic properties for special cases of
$XY$ models with random intersite interactions
were found by
E.R.Smith,$^7$
E.Barouch and B.M.McCoy,$^8$
and R.O.Zaitsev$^9$
using an approach developed by F.J.Dyson.$^{10}$ 
A somewhat different approach was suggested 
by H.Nishimori,$^{11}$
who presented
exact calculations of thermodynamic quantities for the isotropic 
$XY$ model in a random lorentzian transverse field.
Nishimori's exact solution is based on 
the Jordan-Wigner transformation of the spin Hamiltonian 
to a tight-binding model of non-interacting spinless fermions
with diagonal lorentzian 
disorder. 
For the latter fermionic model 
the random-averaged 
one-particle Green functions 
(and hence the density of states that 
yields thermodynamics) 
were found exactly first by P.Lloyd$^{12}$
with the help of contour integrals.
Later on Nishimori's work was 
generalized for models with alternating bonds$^{13}$ 
and 
additional intersite Dzyaloshinskii-Moriya 
interaction.$^{14}$
The main results 
obtained in the papers on models with
a lorentzian transverse field concern the changes in the temperature
dependences of entropy, specific heat and static transverse linear 
susceptibility as well as the ground-state transverse magnetization 
as a function of averaged transverse field once the randomness 
is introduced.

The idea of the present paper is to study the
thermodynamics of a random spin-$\frac{1}{2}$ $XY$ chain 
exploiting an extended version of the Lloyd
model with off-diagonal disorder$^{15}$.
In the late seventies in a series of papers this extended Lloyd model was
applied to disordered electron systems by one of the authors.$^{16-20}$
Similarly to Ref. 11 we present exact calculations
for various thermodynamic
quantities. However, due to the additional off-diagonal disorder our
results concerning the influence 
of disorder on thermodynamic functions differ to some extent from those
obtained by H.Nishimori$^{11}$.

The paper is organized as follows. In Section 2 we describe 
the Jordan-Wigner transformation
from spins to non-interacting spinless fermions and the
evaluation of the averaged one-fermion Green functions.
In Section 3 the averaged Green functions are used to 
calculate the thermodynamic 
properties, namely, entropy, specific heat, transverse magnetization and 
static transverse linear susceptibility.
Some conclusions are 
given in Section 4.
A short report of these results was presented in Ref. 21.

\section{Jordan-Wigner transformation {\protect \\}
         and averaged one-fermion Green functions}

We consider a linear $XY$ chain of $N$ spins 
$\frac{1}{2}$ in a transverse field with periodical boundary conditions. 
The Hamiltonian reads
\begin{eqnarray}
H= \sum_{n=1}^N \Omega_n s_n^z
+\sum_{n=1}^N J_n \left( s^x_ns^x_{n+1}+s^y_ns^y_{n+1}\right) 
\nonumber\\
=\sum_{n=1}^N \Omega_n 
\left(s_n^+s_n^--\frac{1}{2}\right)
+\sum_{n=1}^N 
\frac{J_n}{2} \left( s^+_ns^-_{n+1}+s^-_ns^+_{n+1}\right) ,
\;\;\;s_{n+N}^{\alpha}=s_n^{\alpha},
\end{eqnarray}
where $\Omega_n$ is the transverse field at site $n$ and 
$J_n$ is 
the exchange interaction between the sites $n$ and $n+1$.
The $J_n$ are taken to be independent random variables with 
a lorentzian probability distribution 
\begin{eqnarray}
p(J_n)=\frac{1}{\pi}\frac{\Gamma}
{\left( J_n-J_0\right)^2+\Gamma^2}.
\end{eqnarray}
Here $J_0$ is the mean value and $\Gamma$ is the width of the 
distribution (strength of disorder).
In order to treat the model (1), (2) in an exact manner 
we assume the following relation between
the transverse field at each site $\Omega_n$ and the 
surrounding intersite interactions (cf. Ref. 15-20)
\begin{eqnarray}
\Omega_n-\Omega_0=
a\left( \frac{J_{n-1}-J_0}{2}+ \frac{J_{n}-J_0}{2}\right),
\;\;\; 
a \; {\mbox {is real,}}
\;
\mid a \mid \ge 1, 
\end{eqnarray}
where $\Omega_0$ is the averaged transverse field at site.

Next we transform the spin model to a fermionic model by
Jordan-Wigner transformation: 
$c_1=s_1^-$, 
$c_1^+=s_1^+$, 
$c_j=P_{j-1}s_j^-$,  
$c_j^+=P_{j-1}s_j^+$,
$j=2,...,N$,
$P_j\equiv \prod_{n=1}^j(-2s_n^z)$.  
The resulting Hamiltonian reads
\begin{eqnarray}
H=H^-+BP^+=H^+P^++H^-P^-,
\nonumber\\
H^{\pm}\equiv -\frac{1}{2}\sum_{n=1}^N \Omega_n 
+\sum_{n=1}^N \Omega_n c^+_nc_n
+\sum_{n=1}^N \frac{J_n}{2} \left( c^+_nc_{n+1}-c_nc^+_{n+1}\right) ,
\nonumber\\
B\equiv -J_N\left( c^+_Nc_{1}-c_Nc^+_{1}\right) ,
\nonumber\\
P^{\pm}\equiv \frac{1\pm P}{2}, \;\;\; P\equiv 
\prod_{n=1}^N\left( -2s_n^z\right)
\end{eqnarray}
with anticyclic boundary conditions for $H^+$
and cyclic boundary conditions for $H^-$.
For the calculation of thermodynamic properties of model (1)
one can omit the boundary term $B$,$^{22}$
i.e. it is sufficient to study the 
thermodynamics of spinless
fermions described by 
the $c$-cyclic Hamiltonian $H=H^-$.
This $c$-cyclic fermionic Hamiltonian corresponds to the
one-dimensional version of Anderson's model with
off-diagonal disorder considered by W.John and J.Schreiber.$^{15}$

Following Ref. 15
one introduces the retarded and 
advanced temperature double-ti\-me Green functions 
$G^{\mp}_{nm}(t) \equiv \mp i \theta (\pm t)<\{ c_n(t),c^+_{m}\} >$,
$G^{\mp}_{nm}(t)=\frac{1}{2\pi}
\int_{-\infty}^{\infty}d\omega {\mbox {e}}^{-i\omega t}
G^{\mp}_{nm}(\omega \pm i \varepsilon )$,
$\varepsilon \rightarrow +0$.
For $G^{\mp}_{nm}(\omega \pm i \varepsilon )$ one finds
the following set of equations 
\begin{eqnarray}
\left( \omega \pm i \varepsilon -\Omega_n \right) 
G^{\mp}_{nm}(\omega \pm i \varepsilon ) -
\left[ \frac{J_{n-1}}{2}
G_{n-1,m}^{\mp}(\omega \pm i \varepsilon )+
\frac{J_{n}}{2}
G_{n+1,m}^{\mp}(\omega \pm i \varepsilon ) \right]=
\delta_{nm}.
\end{eqnarray}

Suppose that
$...,J_n,...$ 
(and hence 
$...,\Omega_n,...$)
are complex variables.
The singularities of 
$G^{\mp}_{nm}(\omega \pm i \varepsilon )$
are given by the zeros of
${\mbox {det}}(\omega \pm i \varepsilon -H)$.
${\mbox {det}}(\omega \pm i \varepsilon -H)$ 
is different from zero if the eigenvalues 
$\lambda$ of ${\mbox {Im}}(\omega \pm i \varepsilon -H)$ 
are either all positive or all negative.
>From the Gershgorin criterion$^{23,24}$ for the complex matrix
${\mbox {Im}}(\omega \pm i \varepsilon -H)$
one gets that for any eigenvalue $\lambda$ at least one of the 
conditions
\begin{eqnarray}
\mid {\mbox {Im}}(\omega \pm i \varepsilon 
-\Omega_n)-\lambda \mid
\le \frac{1}{2}\mid {\mbox {Im}}J_{n-1} \mid +
\frac{1}{2}\mid {\mbox {Im}}J_{n} \mid 
\end{eqnarray}
has to be fulfilled.
Using (3) the inequalities (6) can be written as
\begin{eqnarray}
\mid {\mbox {Im}}(\omega \pm i \varepsilon )
-\frac{a}{2}
\left( {\mbox {Im}}J_{n-1}+
{\mbox {Im}}J_{n}
\right)
-\lambda \mid
\le \frac{1}{2}\mid {\mbox {Im}}J_{n-1} \mid +
\frac{1}{2}\mid {\mbox {Im}}J_{n} \mid ,
\;\;\; \mid a \mid \ge 1.
\end{eqnarray}
Let us consider the retarded Green function
(${\mbox {Im}}(\omega + i \varepsilon )>0$).
Then according to (7) 
for $a\ge 1$ all $\lambda$ must be positive
if all ${\mbox {Im}}J_{n}<0$,
whereas for $a\le -1$
all $\lambda$ must be positive
if all ${\mbox {Im}}J_{n}>0$.
Similarly, for the advanced Green function
(${\mbox {Im}}(\omega - i \varepsilon )<0$) 
according to (7)
for $a\ge 1$ all $\lambda$ are negative
if all ${\mbox {Im}}J_{n}>0$,
and for $a\le -1$ all $\lambda$ are negative
if all ${\mbox {Im}}J_{n}<0$.

Consequently, 
for $a\ge 1$ ($a\le 1$)
the retarded Green function
$G_{nm}^-(\omega +i\varepsilon)$
cannot have a pole in the lower (upper) 
half-planes of complex variables $J_n$,
whereas  
the advanced Green function
$G_{nm}^+(\omega -i\varepsilon)$
cannot have a pole in the upper (lower) 
half-planes of complex variables 
$J_n$ for $a\ge 1$ ($a\le -1$).
Using these properties one can perform
the averaging of equations (5),
defined by
\begin{eqnarray}
\overline{(...)} \equiv
\prod_{n=1}^{N}
\int_{-\infty}^{\infty}dJ_n
\frac{1}{\pi}\frac{\Gamma}{\left( J_n-J_0\right)^2+\Gamma^2}(...)
\nonumber\\
=
\prod_{n=1}^{N}
\int_{-\infty}^{\infty}dJ_n
\frac{1}{\pi}\frac{\Gamma}{\left( J_n-J_0+i\Gamma\right) 
\left( J_n-J_0-i\Gamma\right) }(...),
\end{eqnarray}
by means of contour integrals.
For the averaging of a function
$F(...,\Omega_n,J_n,...)$
that has no poles in lower half-planes $J_n$, 
one can close the contours of 
integration in (8) in these half-planes. One obtains
\begin{eqnarray}
\overline{F(...,\Omega_n,J_n,...)} =
F(...,\Omega_0-ia\Gamma,J_0-i\Gamma,...).
\end{eqnarray}
Similarly, for the function without poles 
in upper half-planes $J_n$ one 
gets by contour integration
\begin{eqnarray}
\overline{F(...,\Omega_n,J_n,...)} =
F(...,\Omega_0+ia\Gamma,J_0+i\Gamma,...).
\end{eqnarray}
Then the averaged equations for Green functions (5) 
due to (9), (10) read
\begin{eqnarray}
\left[ \omega \pm i\varepsilon -
\left( \Omega_0\mp i\mid a \mid \Gamma \right) \right]
\overline{G^{\mp}_{nm}(\omega\pm i\epsilon)} -
\nonumber\\
\frac{J_{0}\mp i{\mbox {sgn}}(a) \Gamma}{2}
\left[ \overline{G^{\mp}_{n-1,m}(\omega\pm i\varepsilon)}+
\overline{G^{\mp}_{n+1,m}(\omega\pm i\varepsilon )} \right]=
\delta_{nm}.
\end{eqnarray}

Equations (11) possess translational symmetry 
and therefore they can be 
solved in a standard way.
The resulting averaged Green functions read
\begin{eqnarray}
\overline{G^{\mp}_{nm}(\omega )}=
\frac{1}{2\pi }\int_{-\pi}^{\pi}d\kappa \;
{\mbox{e}}^{i (n-m)\kappa }
\frac{1}
{\omega -
\left[
\Omega_0\mp i\mid a \mid \Gamma+
\left(
J_0\mp i{\mbox {sgn}}(a) \Gamma
\right) \cos{\kappa }
\right]
}
\nonumber\\
=
\frac{\left( \frac{\sqrt{x^2-y^2}-x}{y} \right)^{\mid n-m \mid }}
{\sqrt{x^2-y^2}}
\end{eqnarray}
with
$x \equiv \omega - \Omega_0 \pm i\mid a \mid \Gamma$,
$y \equiv J_0 \mp i {\mbox {sgn}}(a)\Gamma $.

\section{Entropy, specific heat, transverse magnetization and static
         transverse linear susceptibility}

The obtained averaged Green functions (12) 
allow to study thermodynamics 
of the spin model (1)-(3).
For this we diagonalize
the bilinear in Fermi 
operators form $H^-$ (4)
by the canonical transformation
$\eta_k=\sum_{n=1}^Ng_{kn}c_n$,
$\Lambda_kg_{kn}=\sum_{i=1}^Ng_{ki}A_{in}$,
$A_{ij}=\Omega_i\delta_{ij}+
\frac{1}{2}J_i\delta_{j,i+1}+
\frac{1}{2}J_{i-1}\delta_{j,i-1}$,
$\sum_{i=1}^Ng_{ki}g_{pi}=\delta_{kp}$,
$\sum_{p=1}^Ng_{pi}g_{pj}=\delta_{ij}$
with the result
$\sum_{p=1}^N \Lambda_p (\eta^+_p\eta_p - \frac{1}{2})$.
The thermodynamics for certain realization of random intersite interactions
is then determined by the spectrum of
elementary excitations
$\Lambda_p$
or its density
$\rho (E) \equiv \frac{1}{N} \sum_{p=1}^N \delta (E-\Lambda_p)$.
For example, the Helmholtz free 
energy per site is given by
$f=
\frac{1}{N}\{
-\frac{1}{\beta}\ln \prod_{p=1}^N
[\exp (-\frac{\beta \Lambda_p}{2})+
\exp (\frac{\beta \Lambda_p}{2})]\}=
-\frac{1}{\beta}\int dE \rho (E) 
\ln (2{\mbox {ch}}
\frac{\beta E}{2})$.
The result of averaging over realizations of 
random variables is given by
$\overline{f}=
-\frac{1}{\beta}\int dE \overline{\rho (E)}
\ln (2{\mbox {ch}}
\frac{\beta E}{2})$.
The required averaged density of states
is then calculated by 
the averaged one-particle Green functions (12)  
\begin{eqnarray}
\overline{\rho (E)} =
-\frac{1}{\pi }{\mbox {Im}}\overline{G^-_{nn}(E)}=
\frac{1}{\pi }{\mbox {Im}}\overline{G^+_{nn}(E)}
\nonumber\\
=\mp \frac{1}{\pi }{\mbox {Im}}
\frac{1}{ \sqrt{ \left( E- 
\Omega_0 \pm i\mid a \mid \Gamma\right)^2-
\left(J_0 \mp i{\mbox {sgn}}(a)\Gamma\right)^2}}
\nonumber\\
=\frac{1}{\pi}
\sqrt{\frac{\sqrt{A^2+B^2}-A}
{2(A^2+B^2)}},
\nonumber\\
A\equiv (E-\Omega_0)^2+(1-\mid a \mid ^2)\Gamma^2-J_0^2,
\;\;\;
B\equiv 2\Gamma [\mid a \mid (E-\Omega_0)+{\mbox {sgn}}(a)J_0].
\end{eqnarray}
Really,
$\rho (E)=\frac{1}{N}\sum_{p=1}^N \left[
-\frac{1}{\pi}{\mbox {Im}}\Gamma^-_{pp}(E+i\varepsilon )\right] =
\frac{1}{N}\sum_{p=1}^N 
\left[ \frac{1}{\pi}{\mbox {Im}}\Gamma^+_{pp}(E-i\varepsilon )
\right]$,
where
$\Gamma_{pq}^{\mp}(t) \equiv 
\mp i \theta (\pm t)$
$
<\{ \eta_p(t),\eta^+_{q}\} >=
\sum_{i=1}^N\sum_{j=1}^Ng_{pi}g_{qj}
G^{\mp}_{ij}(t)$,
and therefore
$\rho (E)=
\frac{1}{N}\sum_{j=1}^N \left[
-\frac{1}{\pi}{\mbox {Im}}
\right. $
$\left.
G^-_{jj}(E+i\varepsilon )\right] =
\frac{1}{N}\sum_{j=1}^N 
\left[ \frac{1}{\pi}{\mbox {Im}}G^+_{jj}(E-i\varepsilon )
\right]$;
the averaging yields the first two equalities in 
the left-hand side of (13).

Knowing the averaged Helmholtz free energy we can calculate
the entropy and specific heat by the formulae
\begin{eqnarray}
\overline{s}=
\beta^2\frac{\partial \overline{f}}{\partial \beta}=
\int dE \overline{\rho (E)}\left[ \ln{
\left(2{\mbox{ch}}\frac{\beta E}{2}\right)}
-\frac{\beta E}{2}{\mbox{th}}\frac{\beta E}{2}\right],
\\
\overline{c}=
-\beta \frac{\partial \overline{s}}{\partial \beta}=
\int dE \overline{\rho (E)}
\left(
\frac{\frac{\beta E}{2}}
{{\mbox {ch}}{\frac{\beta E}{2}}}\right)^2.
\end{eqnarray}

Due to the magic property of (13)
$\frac{\partial}{\partial \Omega_0}\overline{\rho (E)}=
-\frac{\partial}{\partial E}\overline{\rho (E)}$
one can express transverse magnetization and static transverse
linear susceptibility through the density of states
\begin{eqnarray}
\overline{m_z}\equiv
\overline{<\frac{1}{N}\sum_{n=1}^Ns_n^z>}=
\frac{\partial \overline{f}}{\partial \Omega_0}=
-\frac{1}{2}\int dE \overline{\rho (E)}{\mbox{ th}}\frac{\beta E}{2},
\\
\overline{\chi_{zz}}=
\frac{\partial \overline{m_z}}{\partial \Omega_0}=
-\beta \int dE \overline{\rho (E)}\frac{1}
{(2{\mbox {ch}}{\frac{\beta E}{2}})^2}.
\end{eqnarray}

Let us discuss the obtained results.

Note at first, that in the absence of randomness 
(13) reduces to the well-known result
\begin{eqnarray}
\overline{\rho (E)} =
\mp \frac{1}{\pi }{\mbox {Im}}
\frac{1}{ \sqrt{ \left( E- 
\Omega_0 \pm i\varepsilon \right)^2-
J_0^2}}=
\left\{
\begin{array}{ll}
\frac{1}{\pi} \frac{1}{\sqrt{J_0^2-(E-\Omega_0)^2}}\;\;\; & {\mbox {if}} 
\;\;\;\mid E-\Omega_0\mid \le \mid J_0\mid ,\\ 
0 \;\;\; & {\mbox {otherwise}}
\end{array}
\right.
\end{eqnarray}
as anticipated.

The considered model (1)-(3) essentially differs from
that one with diagonal disorder
treated by H.Nishimori$^{11}$
($J_n=J$, $\Omega_n$ are independent 
random variables with lorentzian 
distributions).
However, Nishimori's model can be obtained 
as a certain limit of the present model, namely
$\Gamma \rightarrow 0$,
$\mid a \mid \Gamma ={\mbox{const}}
=\Gamma_{{\mbox {N}}}$.
The density of states (13) 
in contrast to the case of diagonal disorder is 
not symmetric with respect to the change
$E-\Omega_0 \rightarrow - (E-\Omega_0)$.
However, it remains the same after the replacement
$E-\Omega_0 \rightarrow - (E-\Omega_0)$,
$a \rightarrow -a$,
or 
$E-\Omega_0 \rightarrow - (E-\Omega_0)$,
$J_0 \rightarrow -J_0$,
since the simultaneous change of signs of $J_0$ and $a$
in (13) does not affect $\overline{\rho (E)}$.
Without loss of generality we choose 
$\Omega_0, J_0 > 0$ throughout the rest of the paper.
It is also convenient,
although by no means essential,
to put hereafter $J_0=1$.
The above-mentioned symmetry
of the density of states
can be seen in Figs. 1-3,
where the averaged density of states (13) (Fig. 1),
the averaged density of states (13) in 
comparison with histograms 
$\rho (E)$ 
calculated
for a certain realization of 
random intersite interactions
using exact finite-chain calculations$^{25}$ (Fig. 2),
and
the histograms $\overline{\rho (E)}$ 
obtained by the latter approach for
$\mid a \mid <1$ (Fig. 3) are displayed. The 
density of states for the non-random case (18)
is depicted in Fig. 1 by dashed lines.
For large $\mid a\mid$ 
the edges of the zone are completely smeared out
with increasing strength of disorder $\Gamma$;
for $\mid a\mid \approx 1$
the increase of disorder results in a smearing 
out of mainly one edge of 
the zone.
A further decrease of $\mid a\mid$ up to $0$ 
leads to a recovering of the 
symmetry with respect to
$E-\Omega_0 \rightarrow - (E-\Omega_0)$
and to transforming of
$\overline{\rho (E)}$ 
into 
$\delta (E-\Omega_0)$.
Some consequences induced by this dependence of
$\overline{\rho (E)}$ on $\Gamma$ and $a$
will be seen in the behaviour of thermodynamic quantities.

The results of numerical calculations 
of thermodynamic quantities 
are presented in Figs. 4-13,
namely,
the temperature dependences of the entropy (14) (Figs. 4,5),
the specific heat (15) (Figs. 7,8),
the transverse magnetization (16) (Fig. 11)
and the static transverse linear susceptibility (17) (Fig. 13),
the dependence on averaged transverse field at low temperatures
of the entropy (Fig. 6),
the specific heat (Fig. 9),
the transverse magnetization (Fig. 10)
and the static transverse linear susceptibility (Fig. 12);
the curves that correspond to the non-random case
are depicted in Figs. 4-13 by dashed lines.

The influence of randomness on thermodynamics 
is mainly rather typical.
It leads to 
(i)
a weak deformation of the entropy 
versus temperature curve
with a decrease of the entropy at high temperatures (Figs. 4,5),
(ii)
a broadening and decreasing  
of the peak in the dependence of the specific heat versus 
temperature (Figs. 7,8),
(iii)
a smearing out of the cast in the $\overline{m_z}$ versus
$\Omega_0$ curve at $T=0$ for $\Omega_0=J_0$
and a nonsaturated transverse magnetization at 
any finite transverse field 
(Fig. 10),
(iv)
a decreasing and disappearing of the singularity 
(accompanying the saturation 
of $\overline{m_z}$ 
at $T=0$ for $\Omega_0=J_0$)
in the curve $\overline{\chi_{zz}}$ versus $\Omega_0$ 
at $T=0$ (Fig. 12), 
and (v)
a suppressing of static 
transverse linear susceptibility 
versus temperature curve (Fig. 13). 

However, as can be seen in Figs. 4-13, the 
influence of disorder, especially for small $a$,
essentially depends on the sign of $a$.
Particularly interesting is the case of strong asymmetry
in the density of states 
$\overline{\rho (E)}$
when $\mid a \mid \approx 1$.
>From mathematical point of view the dependence 
of the computed quantities on 
temperature and averaged 
transverse field and the well-pronounced difference 
between the cases $a\approx -1$ and $a\approx 1$
can be understood
having in mind that these quantities according to (14)-(17)
are integrals over $E$ of products of $\overline{\rho (E)}$
(shown in Fig. 1) with functions with 
evident dependence on $E$ at 
different $\beta$.
It is interesting to note that for some 
Hamiltonian parameters and 
temperatures even very large randomness 
(controlled by $\Gamma$) almost
does not affect the thermodynamic quantities.
This can be nicely seen in Figs. 4-13.

It is worth to underline that the asymmetry of 
$\overline{\rho (E)}$
leads to the appearance of nonzero 
averaged transverse magnetization
$\overline{m_z}$ at zero averaged transverse field $\Omega_0$.
As it can be seen from (16) 
$\overline{m_z}=0$ 
at $T=0$, $\Omega_0=0$
if
$\int_{-\infty}^0dE\overline{\rho (E)}=
\int_0^{\infty}dE\overline{\rho (E)}$.
This is evidently true for a symmetric density of states
$\overline{\rho (E)}$
(as in the case considered by H.Nishimori)
but is not obvious in the considered case (13).
The difference between the integrals
$\int_{-\infty}^0dE\overline{\rho (E)}$
and
$\int_0^{\infty}dE\overline{\rho (E)}$
can be clearly demonstrated by numerical 
finite-chain calculations
as a difference between the numbers of negative and positive
eigenvalues $\Lambda_p$ of the $N\times N$ matrix 
$\mid \mid A_{ij} \mid \mid $
denoted by 
${\cal {N}}_-$
and
${\cal {N}}_+$,
respectively.
Examples
for certain realization of the random model (1)-(3)
are given in
Tables I and II.
The transverse magnetization for a certain 
realization at $T=0$ is given by
$m_z =\frac{{\cal {N}}_--{\cal {N}}_+}{2N}$
and one finds a good agreement of 
these direct numerical finite-chain calculations for
$-\overline{m_z}$
with the results depicted in Fig. 10
(e.g. 
for $\Gamma =0.25$
$-\overline{m_z}\approx 0.051, 0.038, 0.011$
if $a=-1.01, -2, -5$ respectively,
for $\Gamma =1$
$-\overline{m_z}\approx 0.095, 0.030, 0.003$
if $a=-1.01, -2, -5$ respectively).
 
\section{Conclusions}

In conclusion, we present exact calculations of the
thermodynamics of the spin-$\frac{1}{2}$
isotropic $XY$ chain with random lorentzian 
intersite interaction and 
a transverse field that depends linearly on the surrounding intersite 
interactions (1)-(3).
The derived exact expressions 
for the averaged density of states (13) and thermodynamic 
quantities (14)-(17) may serve as a testing 
ground for approximate methods 
of spin systems with off-diagonal disorder
that usually involve an unclear error.
Aside from this they are interesting in their own right,
since experimentally accessible systems are 
always affected by randomness,
and an understanding of disorder effects 
even within such simple model
can help in comparing experimental 
observations and theoretical predictions.

Unfortunately, the obtained results do not 
permit to calculate exactly the 
averaged spin correlation functions since 
such calculation requires the 
knowledge of averaged many-particle fermion Green functions.	
Spin correlations and their dynamics may be examined
using exact finite-chain calculations developed in
Refs. 26,27.
 
\section{Acknowledgments}

One of the authors (O.D.) 
would like to thank to T.Krokhmalskii and T.Verkholyak
for helpful discussions.
He
is grateful to 
the Deutscher Akademischer Austauschdienst for 
a scholarship for stay in 
Germany when the present study was started.
He is also indebted to Mr. Joseph Kocowsky for continuous
financial support.
The work was partly 
supported by the Deutsche Forschungsgemeinschaft
(Project Ri 615/1-2).

\clearpage



\noindent
TABLE I.
The numbers of negative and positive $\Lambda_p$,
${\cal {N}}_-$ and
${\cal {N}}_+$,
for three realizations of random intersite interactions
in a finite
model (1)-(3) ($N=1000$)
with $\Omega_0 =0$, $J_0=1$, $\Gamma =0.25$.

\begin{center}

\renewcommand{\arraystretch}{1.5}
\begin{tabular}
{|
c@{\vline}
c@{\vline}c@{\vline}
c@{\vline}c@{\vline}
c@{\vline}c@{\vline}
c@{\vline}c@{\vline}
c@{\vline}c@{\vline}
c@{\vline}c@{\vline}
|}\hline
\multicolumn{1}{|c@{\vline}}{$\;\frac{1}{N}\sum_{j=1}^NJ_j\;$}
& \multicolumn{2}{c|}{$a=-5$}
& \multicolumn{2}{c|}{$a=-2$}
& \multicolumn{2}{c|}{$a=-1.01$}
& \multicolumn{2}{c|}{$a=1.01$}
& \multicolumn{2}{c|}{$a=2$}
& \multicolumn{2}{c|}{$a=5$}\\ \hline
  & ${\cal{N}}_-$ & ${\cal{N}}_+$ 
  & ${\cal{N}}_-$ & ${\cal{N}}_+$ 
  & ${\cal{N}}_-$ & ${\cal{N}}_+$ 
  & ${\cal{N}}_-$ & ${\cal{N}}_+$ 
  & ${\cal{N}}_-$ & ${\cal{N}}_+$ 
  & ${\cal{N}}_-$ & ${\cal{N}}_+$ \\ \hline
0.997 
& $\;\;491\;\;$ & $\;\;509\;\;$  
& $\;\;458\;\;$ & $\;\;542\;\;$ 
& $\;\;454\;\;$ & $\;\;546\;\;$ 
& $\;\;546\;\;$ & $\;\;454\;\;$  
& $\;\;542\;\;$ & $\;\;458\;\;$ 
& $\;\;509\;\;$ & $\;\;491\;\;$ \\ \hline
0.984 & 490 & 510 & 464 & 536 & 442 & 558  
      & 558 & 442 & 536 & 464 & 510 & 490 \\ \hline
1.008 & 487 & 513 & 463 & 537 & 452 & 548  
      & 548 & 452 & 537 & 463 & 513 & 487 \\ \hline
\end{tabular}
\end{center}


\vspace{50mm}



\noindent
TABLE II.
The numbers of negative and positive $\Lambda_p$,
${\cal {N}}_-$
and ${\cal {N}}_+$,
for three realizations of random intersite interactions
in a finite
model (1)-(3) ($N=1000$)
with $\Omega_0 =0$, $J_0=1$, $\Gamma =1$.

\begin{center}

\renewcommand{\arraystretch}{1.5}
\begin{tabular}
{|
c@{\vline}
c@{\vline}c@{\vline}
c@{\vline}c@{\vline}
c@{\vline}c@{\vline}
c@{\vline}c@{\vline}
c@{\vline}c@{\vline}
c@{\vline}c@{\vline}
|}\hline
\multicolumn{1}{|c@{\vline}}{$\;\frac{1}{N}\sum_{j=1}^NJ_j\;$}
& \multicolumn{2}{c|}{$a=-5$}
& \multicolumn{2}{c|}{$a=-2$}
& \multicolumn{2}{c|}{$a=-1.01$}
& \multicolumn{2}{c|}{$a=1.01$}
& \multicolumn{2}{c|}{$a=2$}
& \multicolumn{2}{c|}{$a=5$}\\ \hline
  & ${\cal{N}}_-$ & ${\cal{N}}_+$
  & ${\cal{N}}_-$ & ${\cal{N}}_+$
  & ${\cal{N}}_-$ & ${\cal{N}}_+$
  & ${\cal{N}}_-$ & ${\cal{N}}_+$
  & ${\cal{N}}_-$ & ${\cal{N}}_+$
  & ${\cal{N}}_-$ & ${\cal{N}}_+$ \\ \hline
1.009
& $\;\;495\;\;$ & $\;\;505\;\;$
& $\;\;470\;\;$ & $\;\;530\;\;$
& $\;\;402\;\;$ & $\;\;598\;\;$
& $\;\;598\;\;$ & $\;\;402\;\;$
& $\;\;530\;\;$ & $\;\;470\;\;$
& $\;\;505\;\;$ & $\;\;495\;\;$ \\ \hline
0.986 & 503 & 497 & 471 & 529 & 408 & 592
      & 592 & 408 & 529 & 471 & 497 & 503 \\ \hline
1.034 & 494 & 506 & 469 & 531 & 406 & 594
      & 594 & 406 & 531 & 469 & 506 & 494 \\ \hline
\end{tabular}
\end{center}


\noindent
{\bf List of figure captions}\\
\vspace{0.25cm}

FIG. 1.
The averaged density of states (13)
$\overline{\rho (E)}$ vs. 
$E-\Omega_0$.
\vspace{0.5cm}

FIG. 2.
The averaged density of states (13)
$\overline{\rho (E)}$ 
(broken lines) and the density of states for 
certain realization of random intersite 
interactions obtained by exact 
finite-chain calculations (solid lines) 
vs. $E-\Omega_0$.
\vspace{0.5cm}

FIG. 3.
The density of states 
$\rho (E) $ 
averaged over 10 realizations
for $\mid a \mid<1$ obtained by
exact finite-chain calculations  
vs. $E-\Omega_0$.
\vspace{0.5cm}

FIG. 4.
The entropy $\overline{s}$ (14) vs.
temperature $\frac{1}{\beta}$, 
$\Gamma =0.25$.
\vspace{0.5cm}

FIG. 5.
The entropy $\overline{s}$ (14) vs. 
temperature $\frac{1}{\beta}$, 
$\Gamma =1$.
\vspace{0.5cm}
\noindent

FIG. 6.
The entropy $\overline{s}$ (14) vs. 
transverse field $\Omega_0$,
$\frac{1}{\beta}=0.1$. 
\vspace{0.5cm}

FIG. 7.
The specific heat $\overline{c}$ (15) vs. 
temperature $\frac{1}{\beta}$, 
$\Gamma =0.25$.
\vspace{0.5cm}

FIG. 8.
The specific heat $\overline{c}$ (15) vs.
temperature $\frac{1}{\beta}$, 
$\Gamma =1$.
\vspace{0.5cm}

FIG. 9.
The specific heat $\overline{c}$ (15) vs.
transverse field 
$\Omega_0$,
$\frac{1}{\beta}=0.1$.
\vspace{0.5cm}

FIG. 10.
The transverse magnetization $-\overline{m_z}$ (16) vs.
transverse field 
$\Omega_0$ 
at low temperature ($\frac{1}{\beta}=0.001$).
\vspace{0.5cm}

FIG. 11.
The transverse magnetization $-\overline{m_z}$ (16) vs.
temperature $\frac{1}{\beta}$
at
$\Omega_0=0.5$. 
\vspace{0.5cm}

FIG. 12.
The static transverse linear susceptibility 
$-\overline{\chi_{zz}}$ 
(17) vs. transverse field 
$\Omega_0$ 
at low temperature ($\frac{1}{\beta}=0.001$).
\vspace{0.5cm}

FIG. 13.
The static transverse linear susceptibility 
$-\overline{\chi_{zz}}$ (17) vs.
temperature 
$\frac{1}{\beta}$ 
at $\Omega_0=0.5$.
\vspace{0.5cm}

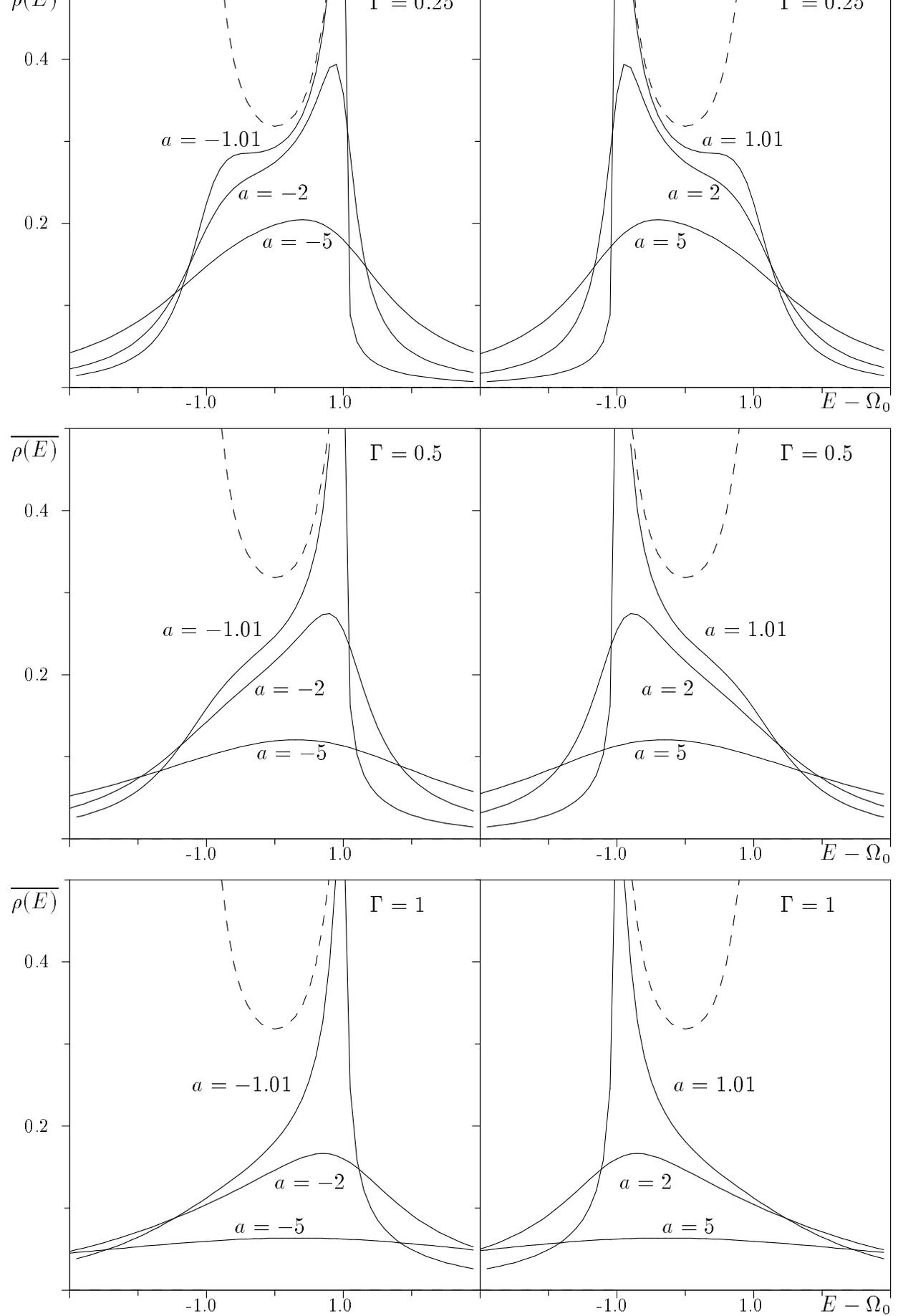
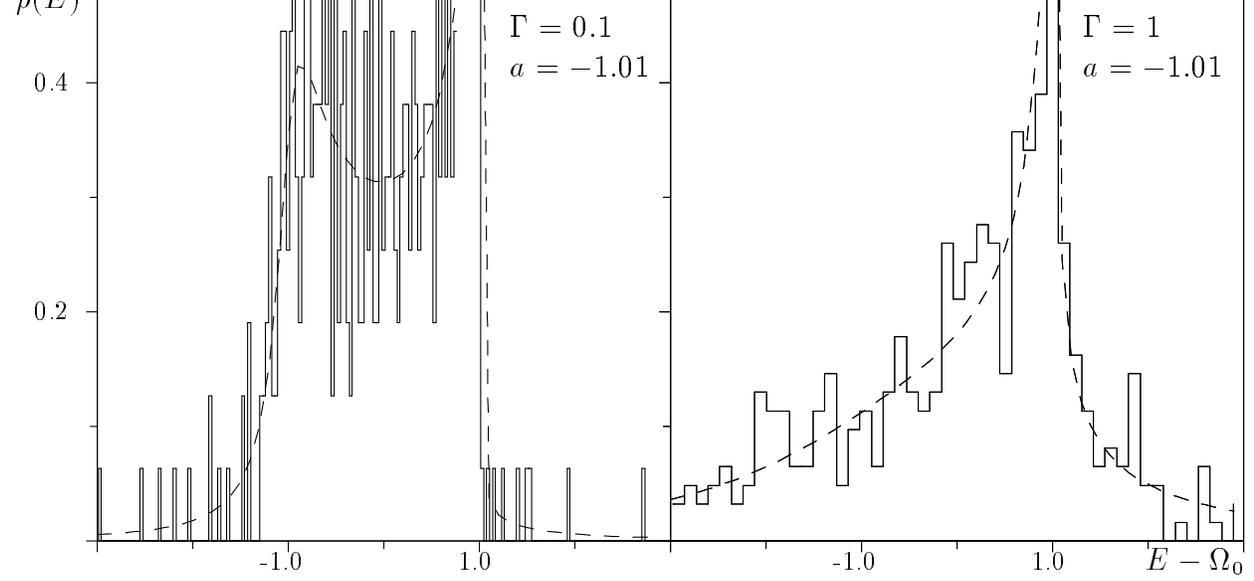
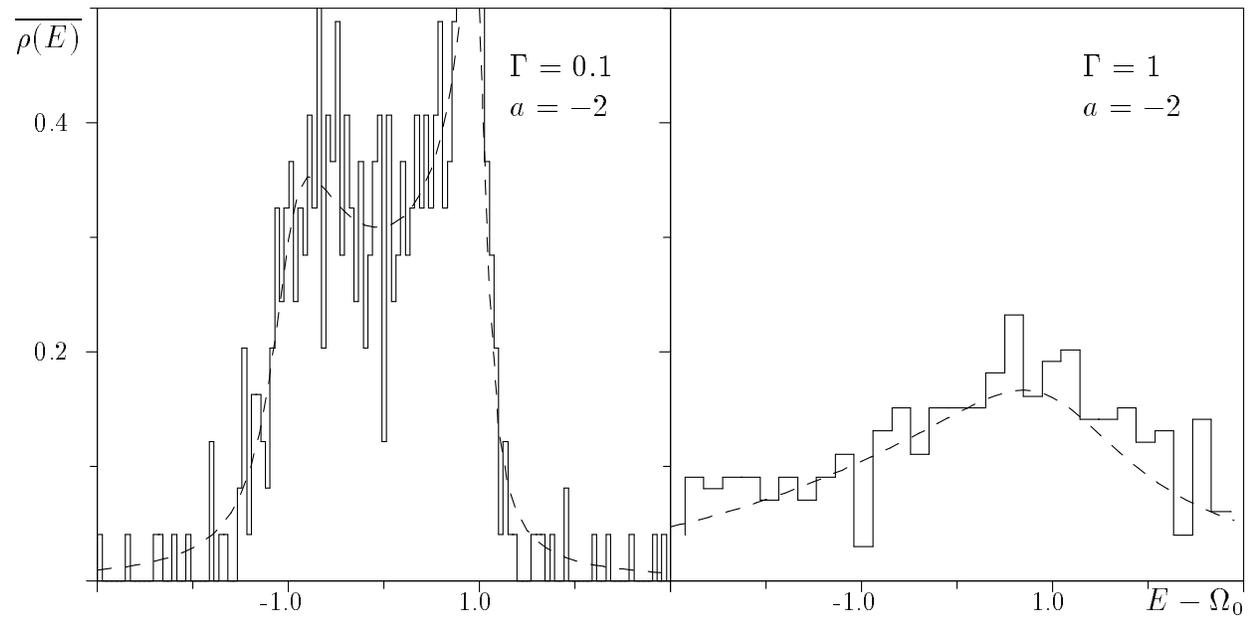
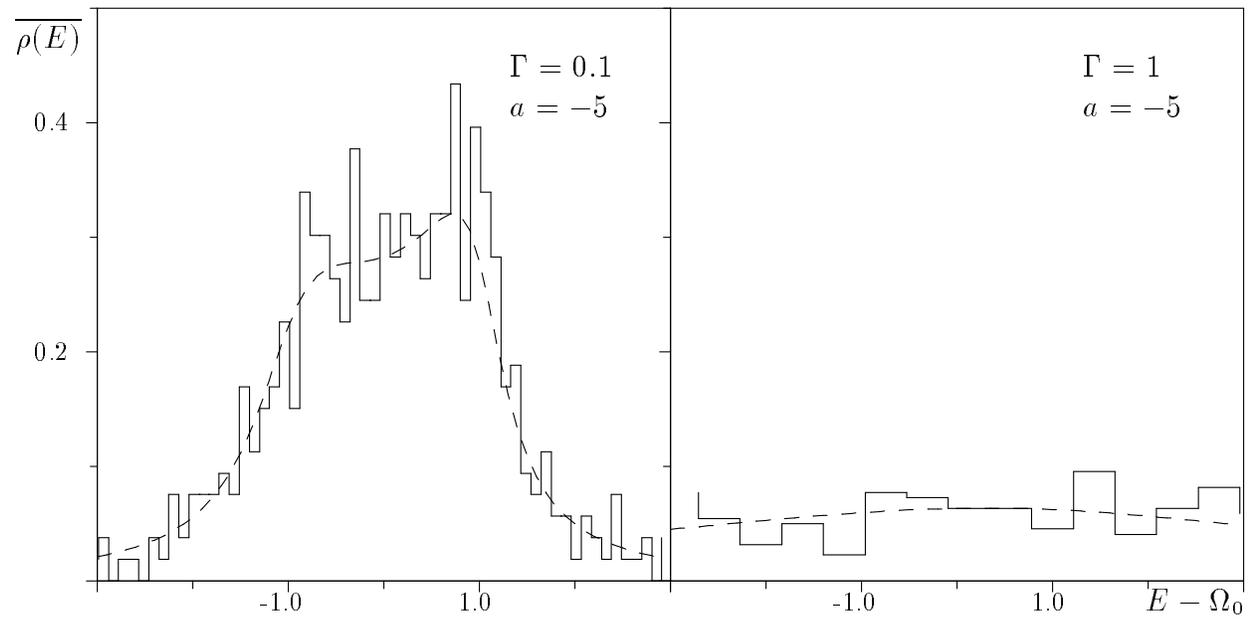
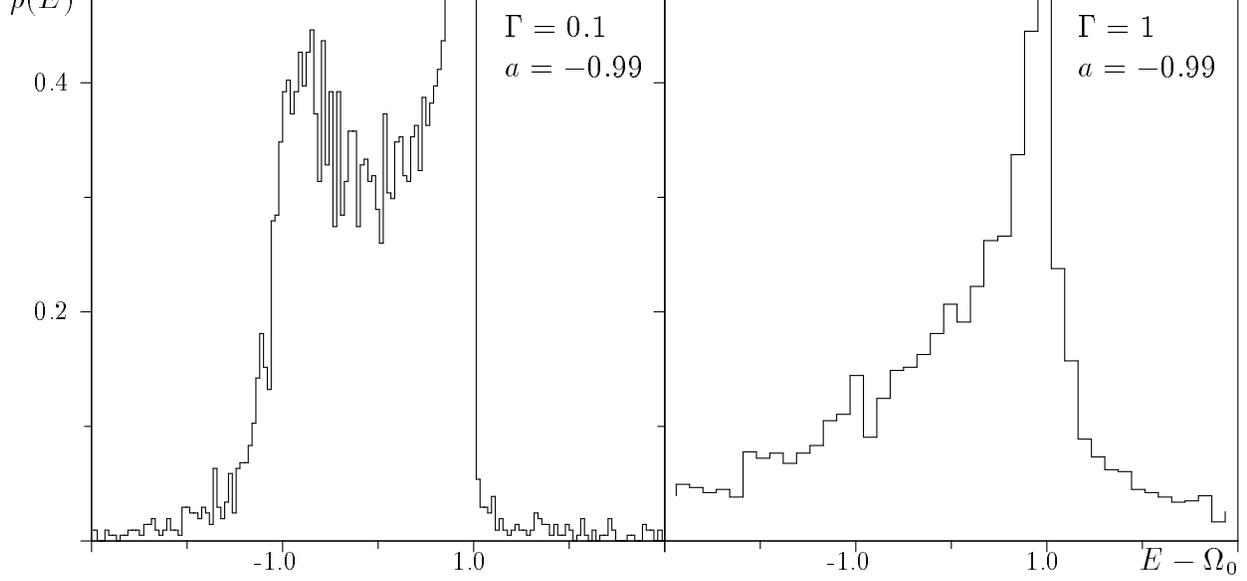
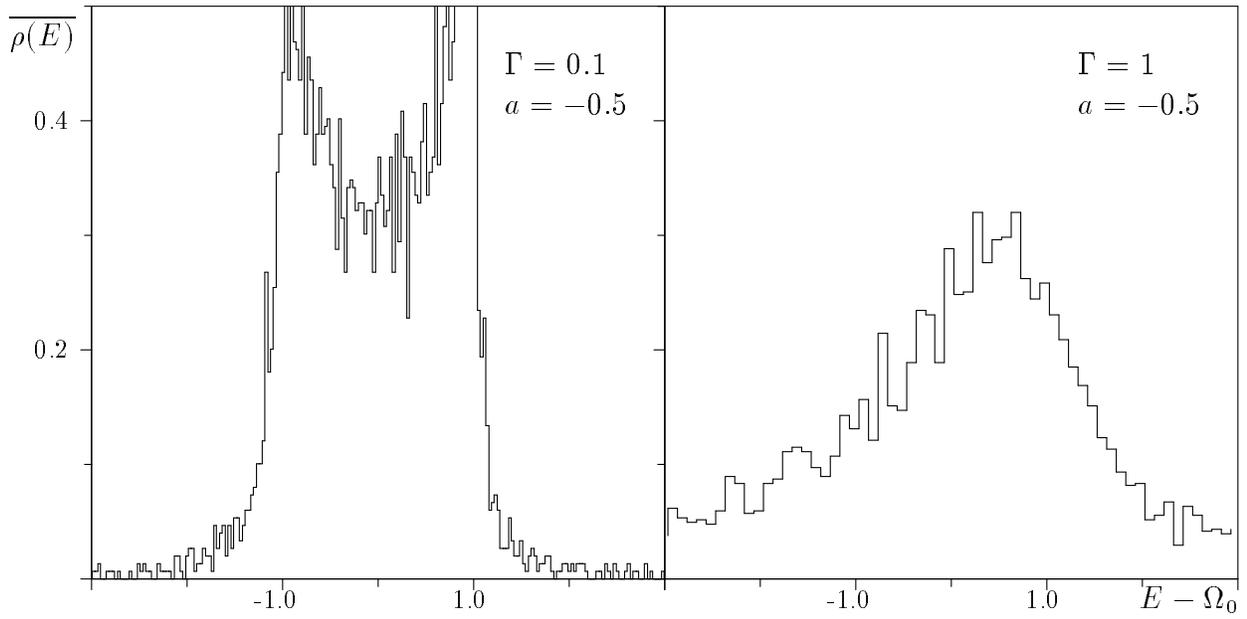
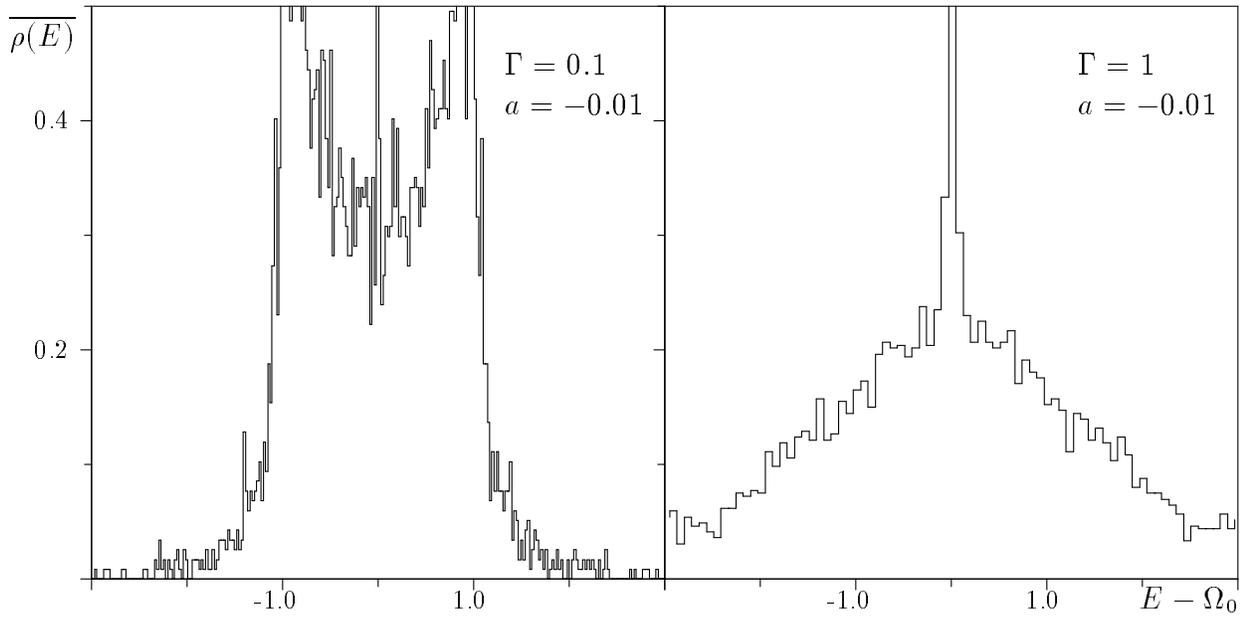
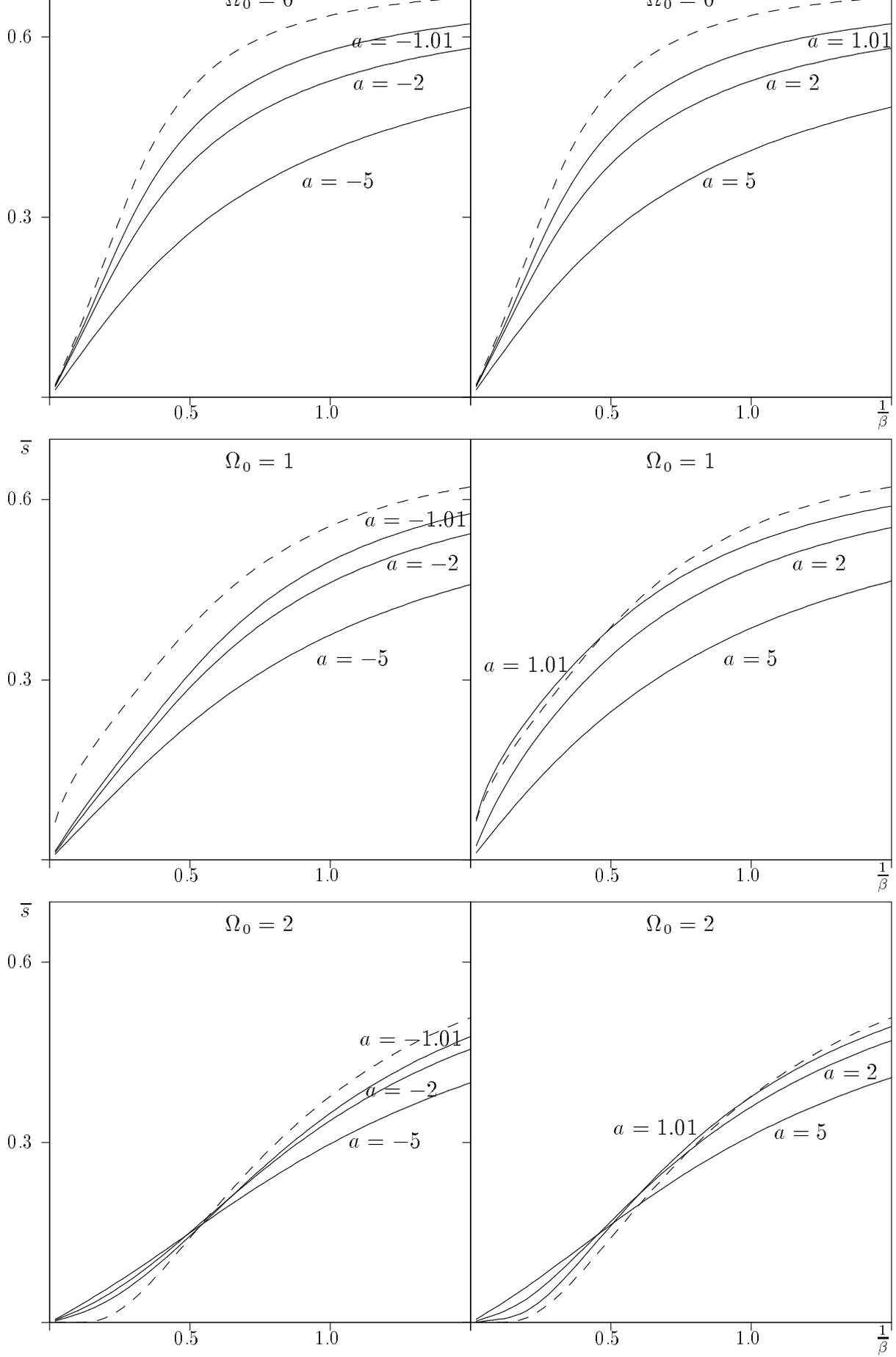
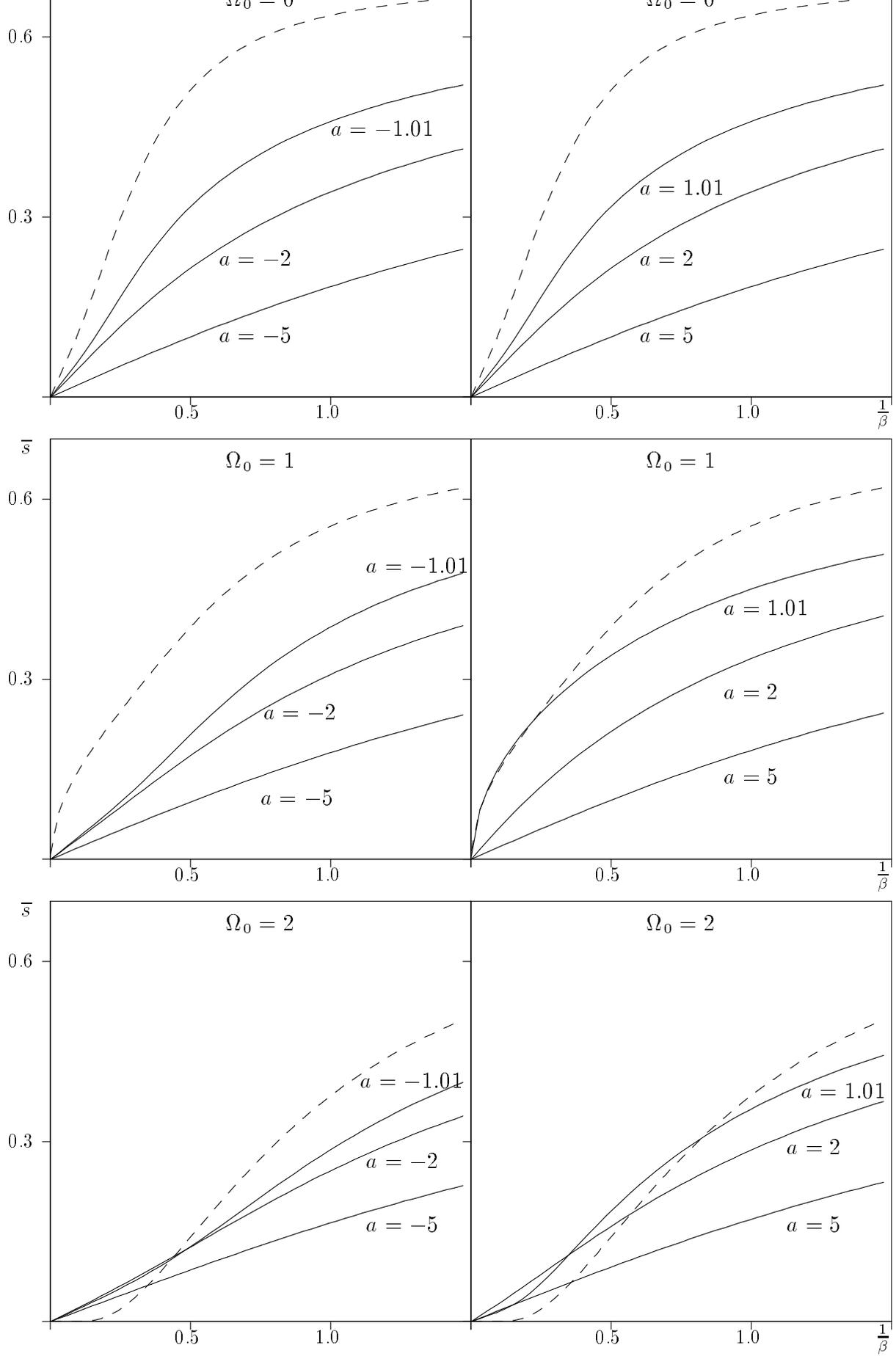
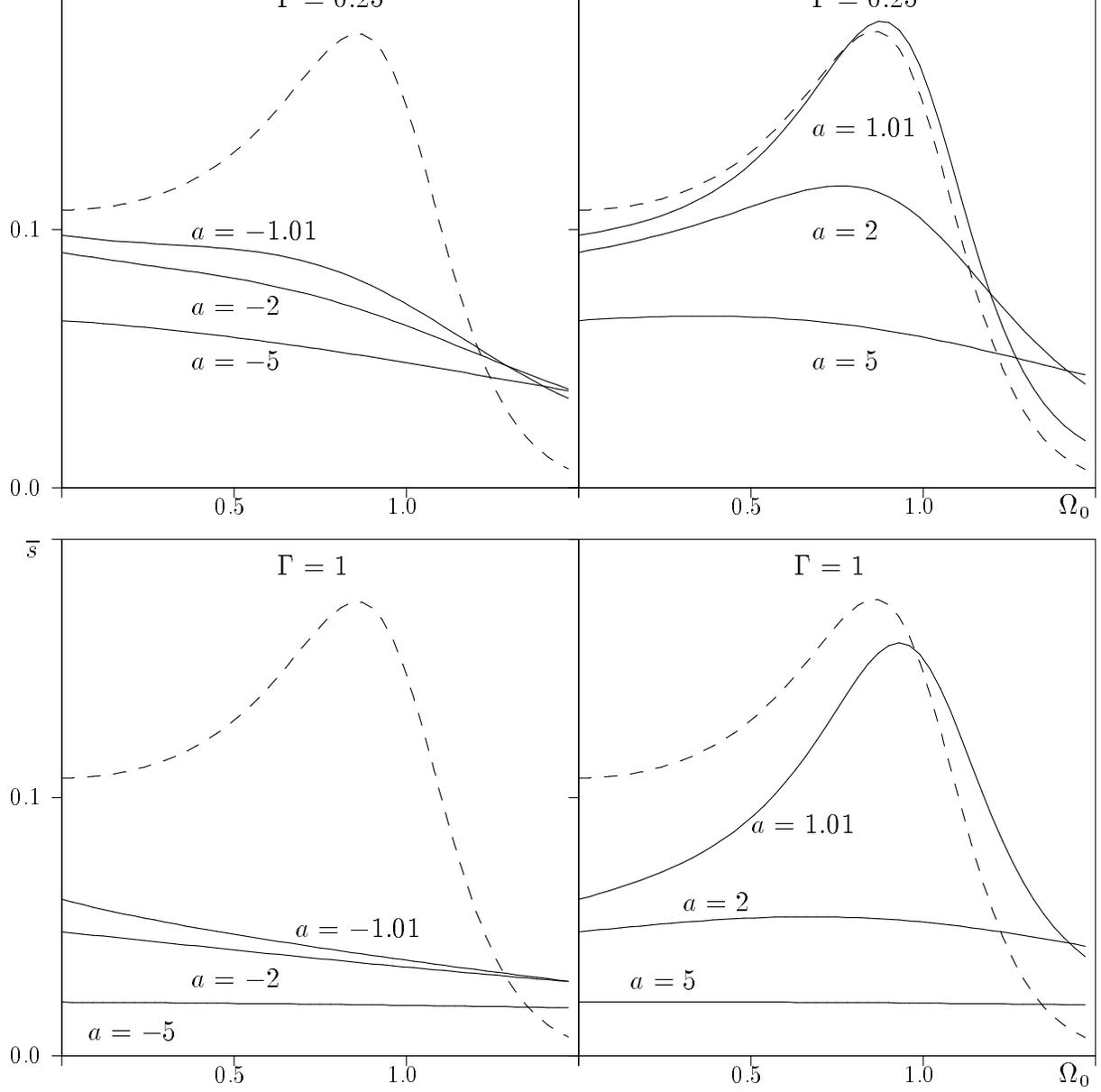
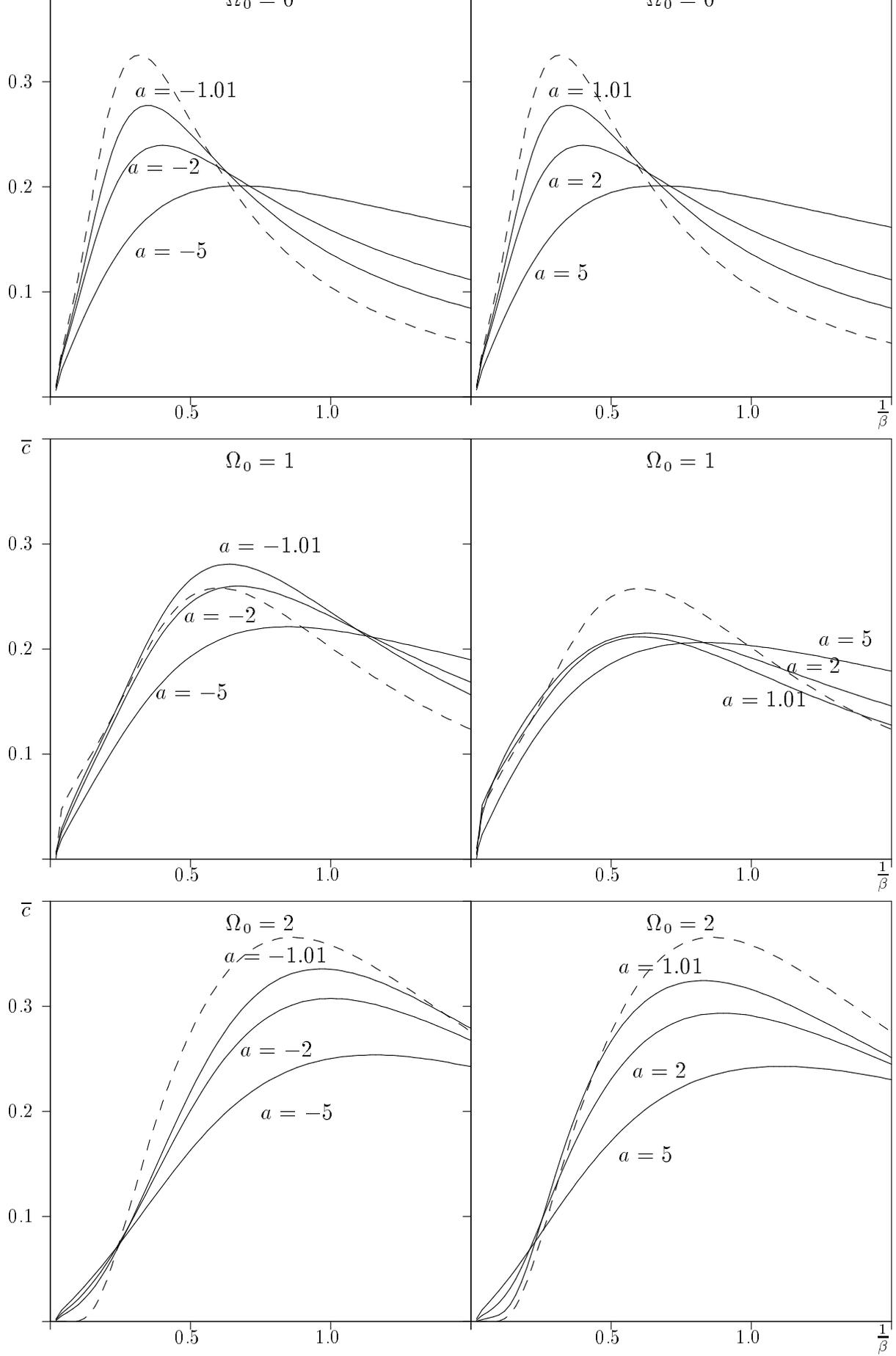
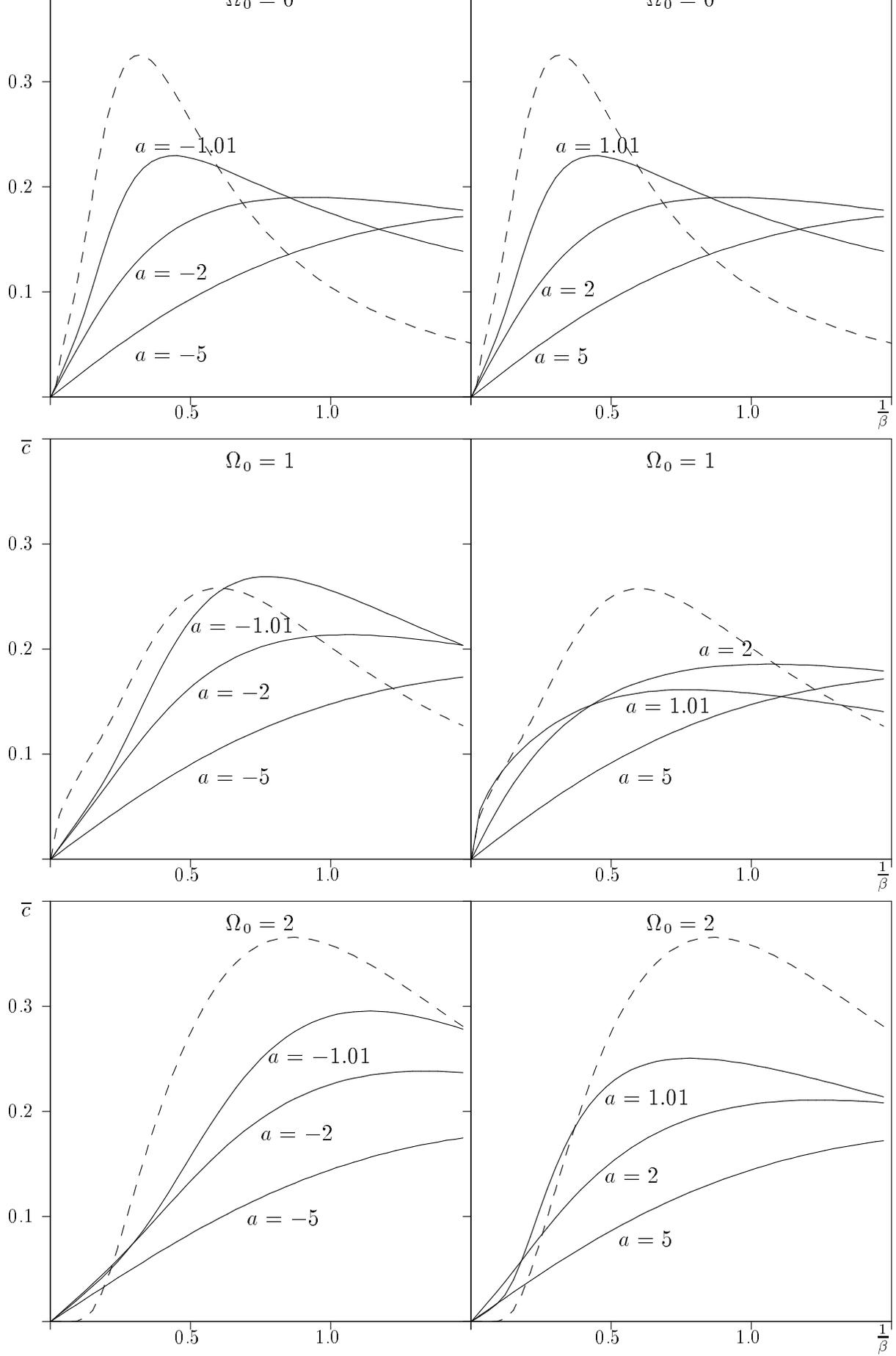
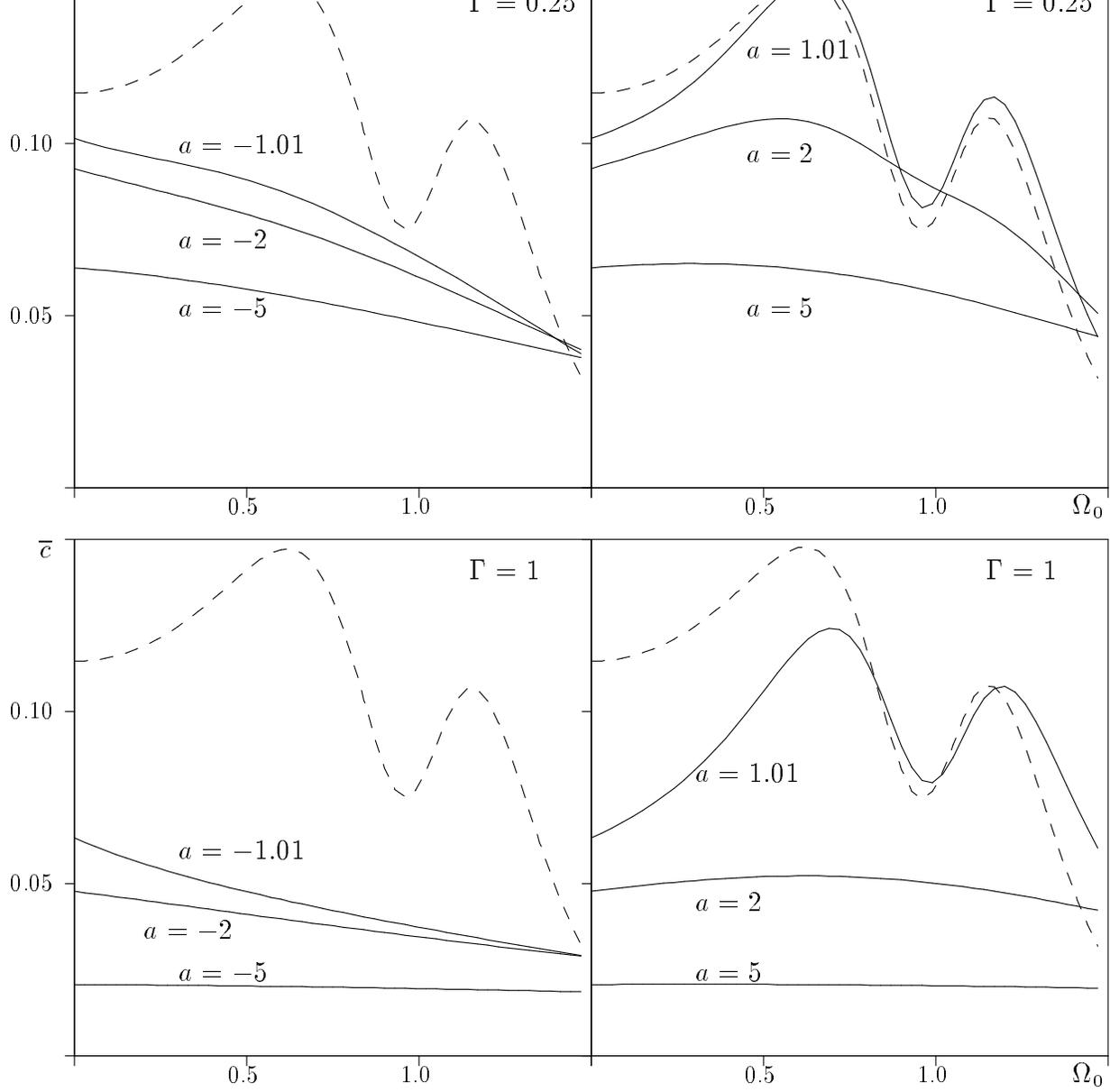
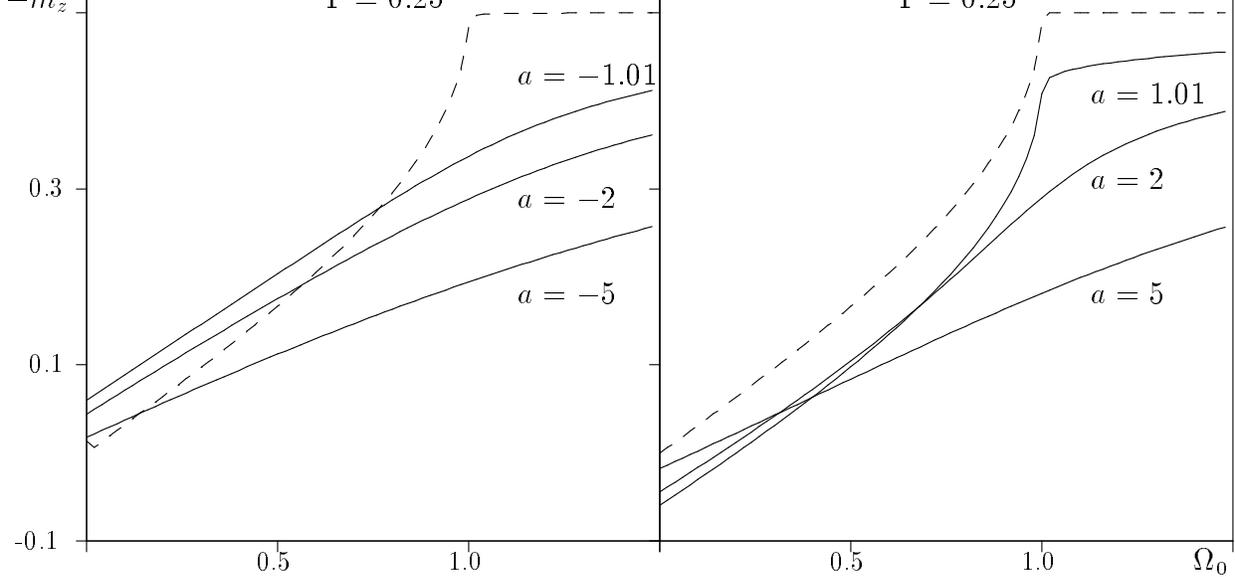
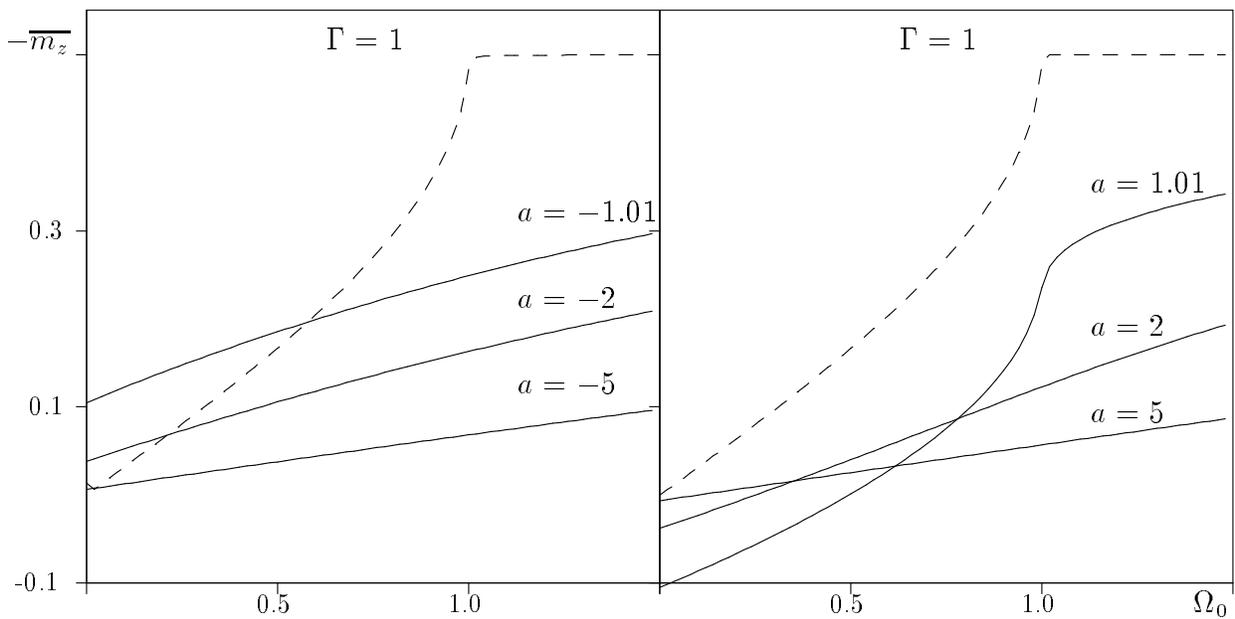
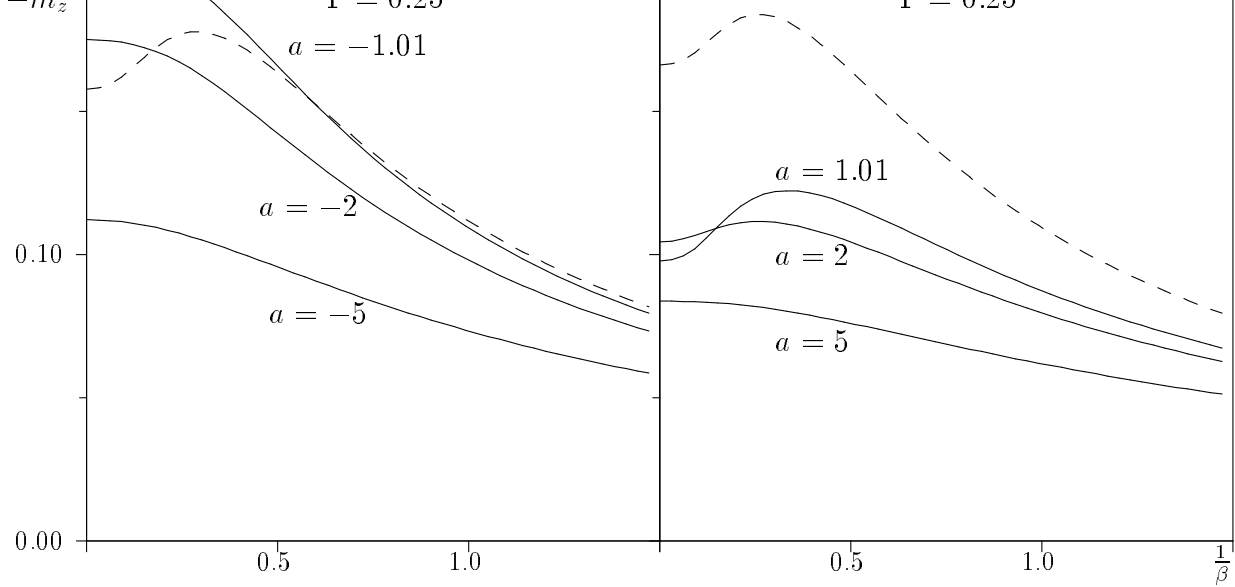
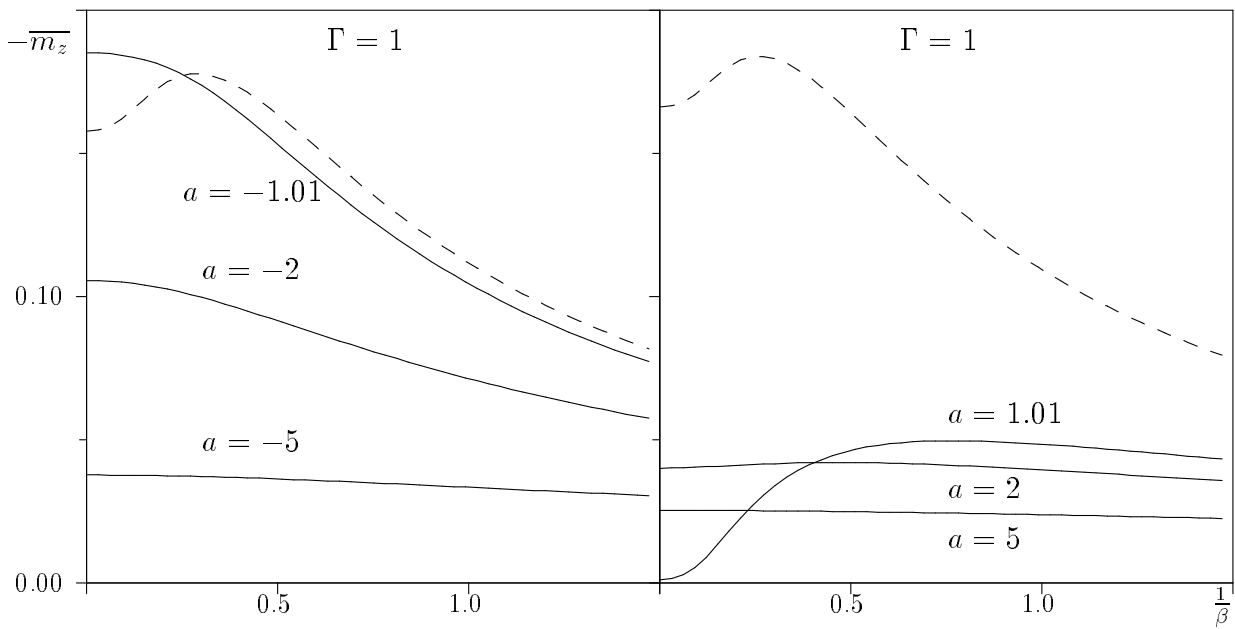
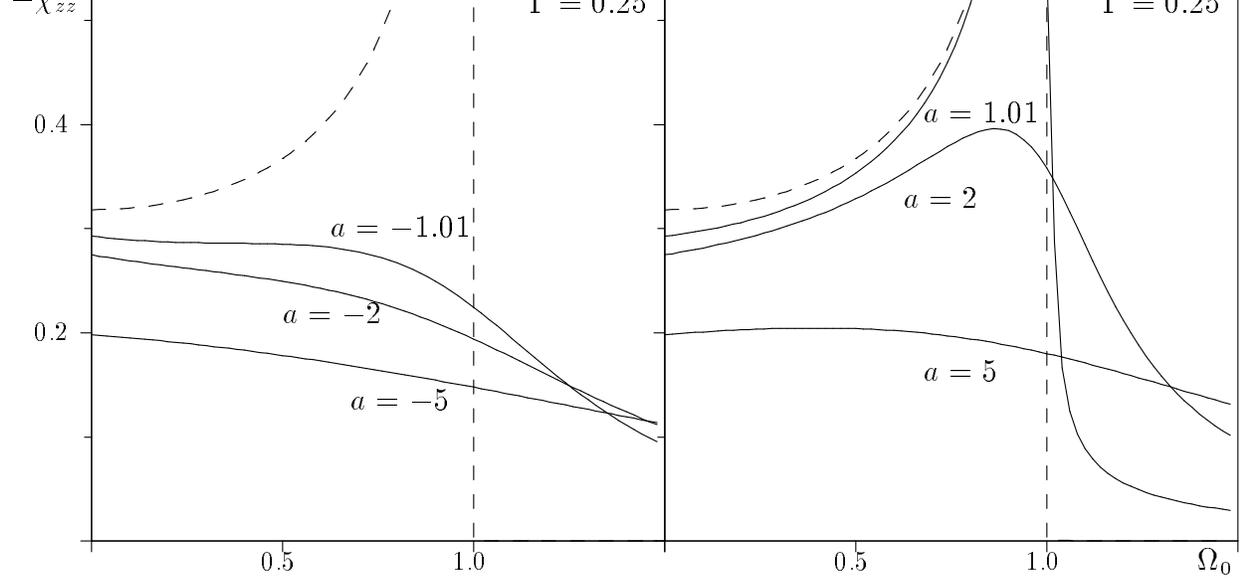
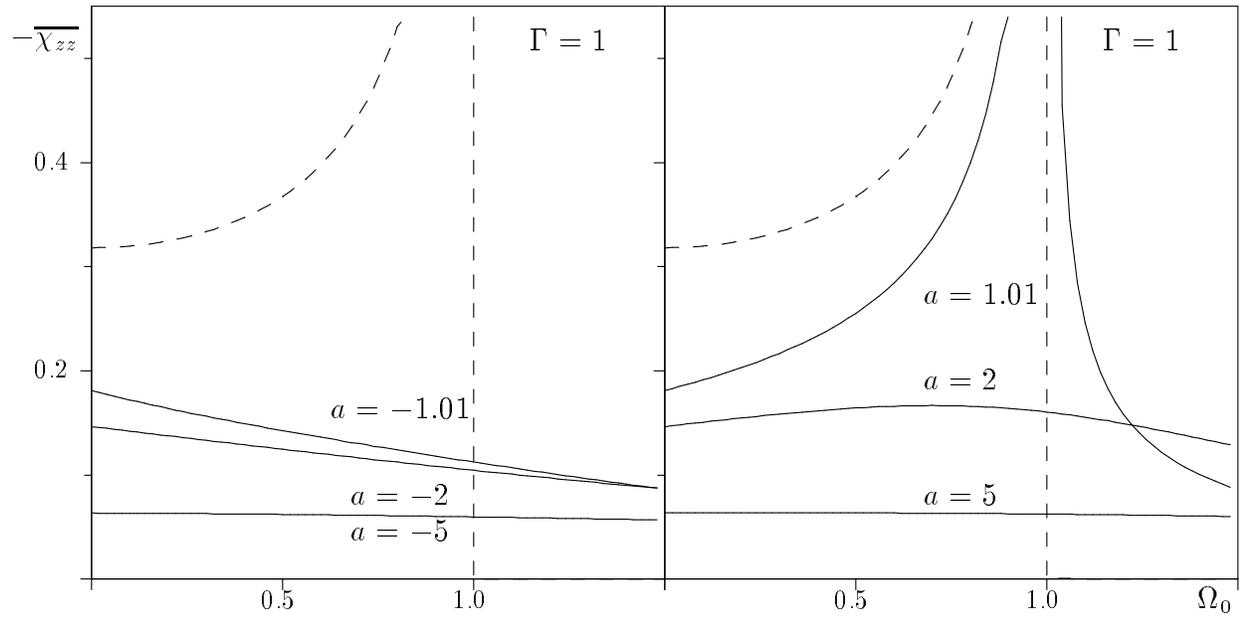
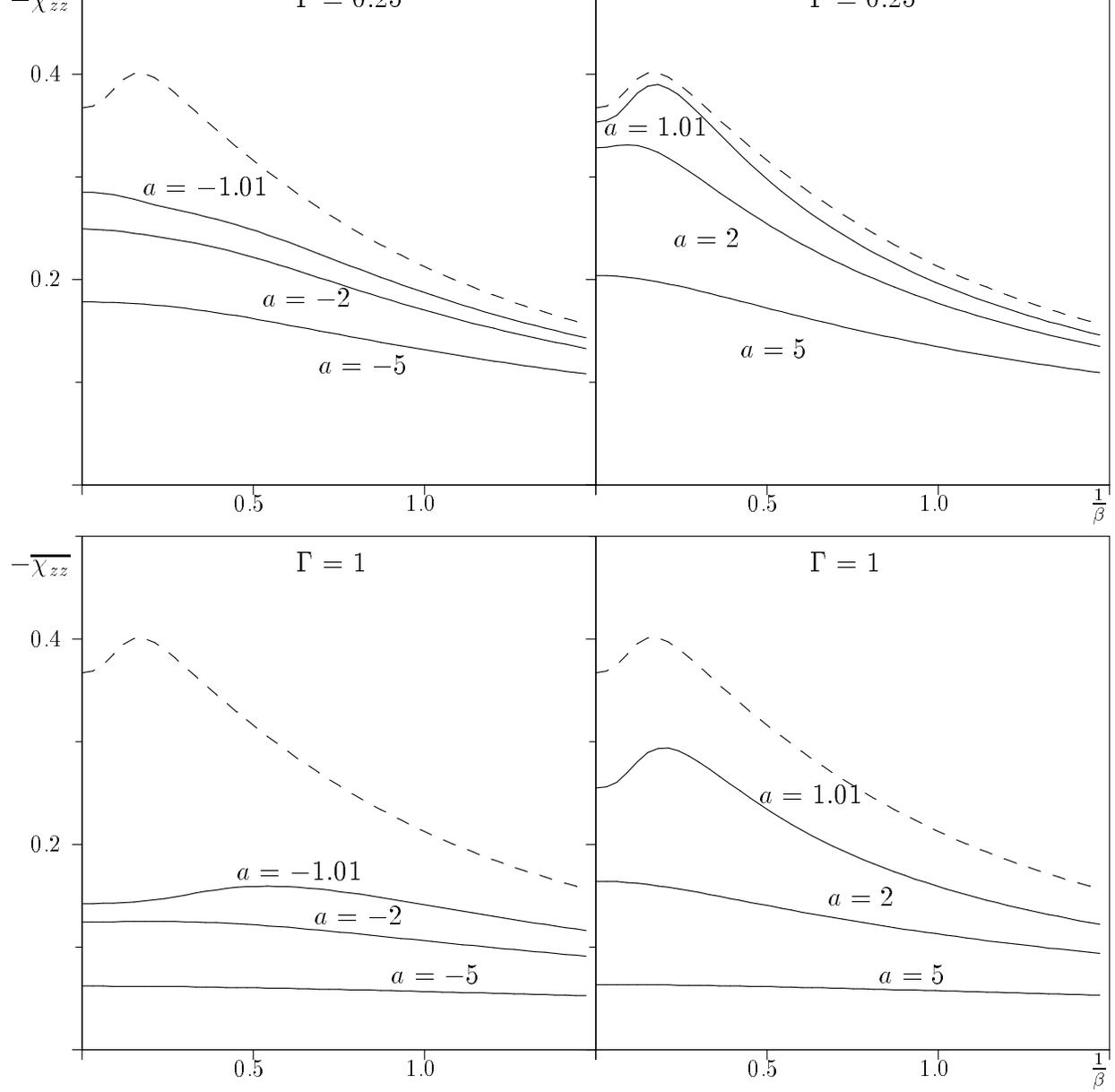
\end{document}